\documentclass[12pt]{iopart}
\usepackage{graphicx, dcolumn, bm, hyperref,color, comment}
\usepackage[square,numbers]{natbib}
\expandafter\let\csname equation*\endcsname\relax
\expandafter\let\csname endequation*\endcsname\relax
\usepackage{amsmath}
\usepackage{mathtools}
\makeatother

\usepackage[normalem]{ulem}

\newcommand{\gae}{\lower 2pt \hbox{$\, \buildrel {\scriptstyle >}\over {\scriptstyle
\sim}\,$}}
\begin{document}
\title[ASEP on a random comb]{Asymmetric simple exclusion process on a random comb: Transport properties in the stationary state}
\author{Mrinal Sarkar$^1$, Shamik Gupta$^2$}
\address{$^1$Institut für Theoretische Physik, Universität Heidelberg, Philosophenweg 19, 69120 Heidelberg, Germany 
}
\address{
$^2$Department of Theoretical Physics,  Tata Institute of Fundamental Research,  Homi Bhabha Road, Mumbai 400005, India
}
\date{\today}
\ead{sarkar@thphys.uni-heidelberg.de,shamik.gupta@theory.tifr.res.in}
\vspace{10pt}
\begin{abstract}
We address the dynamics of interacting particles on a disordered lattice formed by a random comb. The dynamics comprises that of the asymmetric simple exclusion process, whereby motion to nearest-neighour sites that are empty is more likely in the direction of a bias than in the opposite direction. The random comb comprises a backbone lattice from each site of which emanates a branch with a random number of sites. The backbone and the branches run in the direction of the bias. The number of branch sites or alternatively the branch lengths are sampled independently from a common distribution, specifically, an exponential distribution. The system relaxes at long times into a nonequilibrium stationary state. We analyse the stationary-state density of sites across the random comb, and also explore the transport properties, in particular, the stationary-state drift velocity of particles along the backbone. We show that in the stationary state, the density is uniform along the backbone and nonuniform along the branches, decreasing monotonically from the free-end of a branch to its intersection with the backbone. On the other hand, the drift velocity as a function of the bias strength has a non-monotonic dependence, first increasing and then decreasing with increase of bias. However, remarkably, as the particle density increases, the dependence becomes no more non-monotonic. We understand this effect as a consequence of an interplay between biased hopping and hard-core exclusion, whereby sites towards the free end of the branches remain occupied for long times and become effectively non-participatory in the dynamics of the system. This results in an effective reduction of the branch lengths and a motion of the particles that takes place primarily along the backbone.     
\end{abstract}
\maketitle
\section{Introduction}
\label{sec-intro}
Systems driven out of equilibrium exhibit in their stationary state many striking phenomena otherwise absent in equilibrium, i.e., phase transitions and spontaneous symmetry breaking in one dimension with short-range interactions~\cite{evans1998phase,evans1995spontaneous}. Particularly drastic are the effects with quenched disorder in the dynamics, which may be observed even at the level of single-particle dynamics. A representative example is offered by a single particle undergoing biased hopping (i.e., in presence of a field) on a random network, specifically, a random comb (RC), which consists of a one-dimensional backbone lattice from each site of which emanates a branch of random length. The random lengths of the various branches add a source of quenched disorder to the dynamics. The interplay between the quenched disorder and the field in the dynamics leads to various nontrivial phenomena such as a drift velocity that varies non-monotonically with the applied field~\cite{barma1983directed, white1984field, dhar1984diffusion,sarkar2022biased}, anomalous diffusion~\cite{havlin1987anomalous, pottier1995diffusion, bunde1986diffusion, balakrishnan1995transport}. 

Random walks on disordered lattices serve as a useful model for transport in physical systems. In this regard, the RC provides a simple yet nontrivial playground that captures the essential features of many systems with spatial disorder including finitely ramified fractals and percolation clusters~\cite{stauffer1979scaling,rammal1983random,sahimi1993flow}. Dynamics on the RC finds application across fields. Examples include transport in spiny dendrites~\cite{mendez2013comb}, rectification in biological ion channels~\cite{cecchi1996negative}, superdiffusion of ultra-cold atoms~\cite{iomin2012superdiffusive}, reaction-diffusion processes~\cite{agliari2014slow}, crowded-environment diffusion~\cite{benichou2015diffusion}, cancer proliferation~\cite{iomin2006toy}, and human migration along river networks~\cite{campos2006transport}.

In the context of many particles moving in presence of a field, introduction of interactions among the particles enriches the resulting out-of-equilibrium physics, leading to a wide range of complex phenomena. The Asymmetric Simple Exclusion Process (ASEP) serves as a paradigmatic model to understand the out-of-equilibrium physics of interacting many-body systems ~\cite{derrida1998exactly,schutz2001exactly, spitzer1991interaction, liggett2013stochastic}. In the ASEP, the particles are considered indistinguishable and with hard-core interactions. The dynamics involves the particles performing biased random walk between the nearest-neighbour sites of a given lattice. The ASEP finds applications in various domains, including biological transport~\cite{macdonald1968kinetics, parmeggiani2003phase, neri2011totally, neri2013modeling}, pedestrian and traffic flows~\cite{chowdhury2000statistical, antal2000asymmetric, nagatani2001Dynamical}, and quantum dot transport~\cite{ono2002current}. Besides, the ASEP provides a theoretical framework for understanding various physical phenomena in the domain of nonequilibrium physics, such as cluster dynamics~\cite{pronina2004two}, spontaneous symmetry breaking~\cite{evans1995spontaneous}, domain-wall dynamics~\cite{krug1991boundary, janowsky1992finite, popkov2003localization,hinsch2006bulk}, phase separation~\cite{evans1998phase, kafri2002criterion}, and boundary-induced phase transitions~\cite{blythe2007nonequilibrium}. 

Our present work is a revisit of the problem of the ASEP dynamics taking place on the RC lattice~\cite{ramaswamy1987transport}. Earlier studies have considered the ASEP dynamics on an infinite cluster in the percolation above the percolation threshold~\cite{ramaswamy1987transport}. It was shown that the stationary-state drift velocity is always non-zero, exhibiting non-monotonicities as functions of the field and the ASEP particle density. A recent work addressed the stationary-state probability distribution  of the waiting time $T_w$ of a randomly-chosen particle in a side-branch since its last step along the backbone~\cite{iyer2025asymmetric}. 

The ASEP dynamics settles at long times into a nonequilibrium stationary state (NESS) for which obtaining exact results has been a long-standing problem~\cite{golinelli2006asymmetric}.
Most studies on the ASEP have been for the one-dimensional lattice, with only a few works on networks~\cite{neri2011totally,brankov2004totally, embley2009understanding,neri2013exclusion,Mottishaw_2013}. Exact analytical results obtained so far have been for one-dimensional lattice~\cite{derrida1993exact}, and only very recently, results in higher integer dimensions have been obtained~\cite{ishiguro2024exact}. The RC that we study represents a spatially-disordered system with spectral dimension $d_s$ lying in the range $1<d_s<2$. 

In this work, we obtain the exact stationary-state density profile for the ASEP dynamics on the RC. By mapping the problem to an ASEP on a periodic ring, we also compute the drift velocity of the particles along the backbone in the stationary state, a topic that has been studied actively in the past, see, e.g., Ref.~\cite{ramaswamy1987transport}; we discuss the nontrivial interplay between interactions, quenched disorder and field in dictating the behavior of the stationary-state drift velocity. Further, we validate our theoretical predictions via extensive Monte Carlo simulations.  

The paper is organized as follows: Sec.~\ref{sec:model} defines the ASEP dynamics and the random comb in detail. Section~\ref{sec:asep_comb} presents the main results: the exact stationary-state site densities and drift velocity along the backbone. Finally, Sec.~\ref{sec:summary} summarizes our findings and outlines potential future directions. Some technical details as well as details of Monte-Carlo simulations of the studied dynamics are provided in the two appendices.

\section{Model and dynamics}
\label{sec:model}
\begin{figure}[!ht]
	\centering	\includegraphics[scale=0.5]{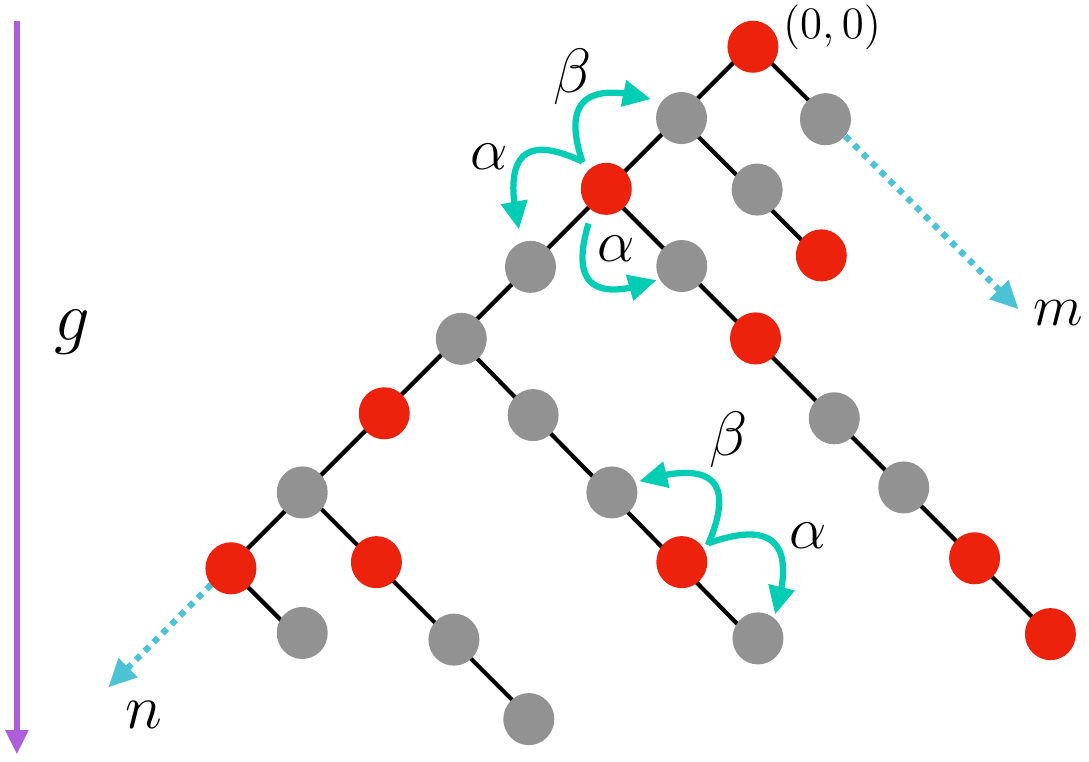}
	\caption{A schematic of the random comb, consisting of a backbone, with random-length branches. Here, $g$ denotes a constant field acting downward, while the quantities $\alpha$ and $\beta$ denote rates with which an ASEP particle hops to a nearest-enighbor site along and opposite to the direction of the field, respectively. The pair of indices ($n,m$) label the sites on the comb, wherein $0\leq n \leq N-1$ labels the backbone sites and $0 \leq m \leq L_{n}$ labels the $(L_n+1)$ number of sites on the branch attached to the $n$-th backbone site. Presence and absence of red filled circle denote respectively an occupied and an empty site.}
	\label{fig:comb}
\end{figure}

We now discuss our model in more detail. The RC-backbone (Fig.~\ref{fig:comb}) is a one-dimensional ($1D$) lattice of $N$ sites, to each of which is attached a branch of a $1D$ lattice with a random number of sites. All lattice spacings in the RC are taken to be unity. We allow the branch lengths to have a maximum-allowed value that we denote by $M$. Moreover, we denote by the pair of indices ($n,m$) the sites on the comb, wherein $0\leq n \leq N-1$ labels the backbone sites and $0 \leq m \leq L_{n}$ labels the $(L_n+1)$ number of sites on the branch attached to the $n$-th backbone site. The site ($n,m=0$) being shared by both the backbone and the branch, we will from now on refer to branch sites as those that have $m>0$. The $L_{n}$'s are quenched-disordered random variables drawn independently from an arbitrary distribution $\mathcal{P}_L$.  The backbone and branches run along a field or a bias with strength $g$; $0 < g < 1$. We will consider for $\mathcal{P}_L$ the representative choice of an exponential:
\begin{align}
\mathcal{P}_L= \frac{{1 - e^{- {1}/{\xi}}} }{1-e^{- (M+1)/\xi}} e^{- L/\xi};~0 \le L \le M.
	\label{eq:exponential}
\end{align}
The ASEP evolution in an infinitesimal time interval $[t,t+\mathrm{d}t]$ involves a particle on a site attempting biased hopping: hop to nearest-neighbor site(s) along (respectively, against) the bias with rate $\alpha \equiv W(1 + g)$ (respectively, $\beta \equiv W(1-g)$). Hard-core exclusion implies that the attempt in each case succeeds only if the destination site is unoccupied before the attempted hop. We assume respectively periodic and reflecting boundary conditions for the backbone and the open end of each of the branches, and define a quantity $f \equiv \alpha/\beta >1$ for later use. 

The ASEP system will be defined by a given total number of particles, which will evidently be conserved by the dynamics on the RC. Note that every realization of the RC has in general a different total number of lattice sites, and consequently, owing to hard-core exclusion, the ASEP on a given RC realization can have a maximum total particle number given by the corresponding total number of lattice sites. For a given realization of the RC, we will use the notation $\rho_\mathrm{tot}$ as the total ASEP particle density, given by the ratio of the total number $\mathcal{N}$ of ASEP particles whose dynamics on the RC one has chosen to study and the total number of sites comprising the RC, given by $\mathcal{L}\equiv \sum_{n=0}^{N-1}(L_n+1)$. It is evident from the dynamics that for a given realization of the RC and a given value of $\mathcal{N}$, the system is ergodic: any configuration of particles on the RC can be reached from any other through the dynamical rules of evolution. An immediate consequence is that the NESS that system settles into at long times is unique. 

 \section{The exact stationary state}
\label{sec:asep_comb}

Let $\mathrm{n}_{n,m}=0/1$ denote the occupation of the $(n,m)$-th site on the RC. The probability $P(C,t)$ of observing a given configuration $C=\{\mathrm{n}_{n,m}\}$ at time $t$ evolves following the master equation 

\begin{align}
    \frac{\partial P(C, t)}{\partial t} = \sum_{C' \neq C} \left[ W(C' \to C) P(C', t) - W(C \to C') P(C, t) \right].
    \label{eq:ME-RC}
\end{align}
Here, the first term in the summation represents probability inflow into $C$, while the second term represents probability outflow from $C$. The quantity $W(C' \to C)$ is the transition rate from configuration $C'$ to $C$; we have $W(C' \to C) = \alpha$ or $\beta$ as set by the dynamical rules of evolution. The explicit form of the master equation is provided in \ref{app0}.

The stationary state has the left hand side of Eq.~\eqref{eq:ME-RC} equal to zero; we denote the corresponding time-independent probability by $P(\{\mathrm{n}_{n,m}\})$. The stationary-state single-site density is given by
\begin{align}
    \rho_{n,m}=\langle \mathrm{n}_{n,m} \rangle=\sum_{\{\mathrm{n}_{n,m}\}}\mathrm{n}_{n,m} P(\{\mathrm{n}_{n,m}\}).
    \label{eq:rho_def}
\end{align}
Multiplying the stationary-state form of Eq.~\eqref{eq:ME-RC} by $\mathrm{n}_{n,m}$ and then summing over $\mathrm{n}_{n,m}=0,1$, we get for the branch sites (for all $n$)
\numparts
\begin{align}
    0&= \alpha \langle \mathrm{n}_{n,m-1}(1-\mathrm{n}_{n,m})\rangle  + \beta \langle (1-\mathrm{n}_{n,m})\mathrm{n}_{n,m+1}\rangle \nonumber\\ 
    &\quad - \alpha \langle \mathrm{n}_{n,m} (1-\mathrm{n}_{n,m+1})\rangle 
    -\beta  \langle (1-\mathrm{n}_{n,m-1})\mathrm{n}_{n,m}\rangle;\qquad 0<m<L_n
    \label{eq:branch-again-0}\\
    0&= \alpha \langle \mathrm{n}_{n,L_n-1} (1 - \mathrm{n}_{n, L_n}) \rangle - \beta \langle \mathrm{n}_{n, L_n} (1 - \mathrm{n}_{n,L_n-1}) \rangle; \qquad m=L_n,
    \label{eq:branch-again-1}
\end{align}
\endnumparts
while for the backbone sites ($m=0$), we have 
\begin{align}
    0&= \alpha \langle \mathrm{n}_{n-1,0}(1-\mathrm{n}_{n,0})\rangle + \beta \langle (1-\mathrm{n}_{n,0})\mathrm{n}_{n+1,0}\rangle - \alpha \langle \mathrm{n}_{n,0}(1-\mathrm{n}_{n+1,0})\rangle \nonumber\\
    &\quad -\beta \langle (1-\mathrm{n}_{n-1,0})\mathrm{n}_{n,0}\rangle - \alpha \langle \mathrm{n}_{n,0}(1-\mathrm{n}_{n,1})\rangle 
    +\beta \langle (1-\mathrm{n}_{n,0})\mathrm{n}_{n,1}\rangle; \quad 0 \leq n \leq N-1.\label{eq:backbone-again-0}
\end{align}
Note that periodic boundary condition implies that $\mathrm{n}_{N,0} = \mathrm{n}_{0,0}$ and $\mathrm{n}_{-1,0} = \mathrm{n}_{N-1,0}$.

For the ASEP on a periodic lattice, one has in the context of the master equation~\eqref{eq:ME-RC} that the stationary probability $P(C)$ satisfies pairwise balance $P(C)W(C \to C')=P(C'')W(C'' \to C)$ for $C' \ne C''$, with $P(C)$ having a product measure form: a product of factors for individual sites, with the factor having the same form for all the sites~\cite{Blythe_2007}; a consequence is that the stationary-state single-site density is uniform across the lattice. A generalization of the product-measure result to our model would imply the stationary state having the form
\begin{align}
P(\{\mathrm{n}_{n,m}\})=\frac{\prod_{n,m} f_{n,m}(\mathrm{n}_{n,m})\delta_{\sum_{n,m}\mathrm{n}_{n,m},\mathcal{N}}}{Z(\mathcal{L},\mathcal{N})},
\label{eq:product-measure-RC}
\end{align}
where now the factors $f_{n,m}(\mathrm{n}_{n,m})$ for individual sites vary from site to site, while the delta-function constraint enforces the total particle number in the system to equal the given number $\mathcal{N}$. A detailed discussion of the stationary-state measure on the RC is given in~\ref{app2}. In the above equation, the factor $Z(\mathcal{L},\mathcal{N})$ plays the role of the normalization constant, and is given by
\begin{align}
    Z(\mathcal{L}, \mathcal{N}) = \sum_{\{\mathrm{n}_{n,m} \} }\prod_{n,m} f_{n,m}(\mathrm{n}_{n,m})\delta_{\sum_{n,m} \mathrm{n}_{n,m},\mathcal{N}}=\frac{1}{2\pi i} \oint dz \, z^{-(\mathcal{N}+1)} \prod_{n,m} F_{n,m}(z),
    \label{eq:ZLN}
\end{align}
where in the last step, we have used the integral representation of the delta function, and have defined the function $F_{n,m}(z)$ as
\begin{align}
    F_{n,m}(z) \equiv \sum_{\mathrm{n}_{n,m}} f_{n,m}(\mathrm{n}_{n,m}) z^{\mathrm{n}_{n,m}}.
\end{align}

In the thermodynamic limit $\mathcal{L} \to \infty$, $\mathcal{N} \to \infty$, keeping the ratio $\mathcal{N}/\mathcal{L}$ fixed and finite and equal to the ASEP particle density $\rho_\mathrm{tot}$, the integral for $Z(\mathcal{L}, \mathcal{N})$ in Eq.~\eqref{eq:ZLN} may be evaluated by saddle-point approximation. The equation defining the saddle point yields 
\begin{align}
\rho_\mathrm{tot} = \frac{z}{\mathcal{L}} \sum_{n,m} \frac{\partial}{\partial z} \ln F_{n,m}(z).
\label{eq:ASEP-rho}
\end{align}
For given $\rho_\mathrm{tot}$, the above equation fixes the value of the quantity $z$. Since we have in any given configuration $\{\mathrm{n}_{n,m}\}$ that $\mathcal{N}=\sum_{n,m}\mathrm{n}_{n,m}$, we find on averaging that $\rho_\mathrm{tot}=(1/\mathcal{L})\sum_{n,m}\rho_{n,m}$. On comparison with Eq.~\eqref{eq:ASEP-rho}, we get 
\begin{align}
  \rho_{n,m} = z \frac{\partial}{\partial z} \ln F_{n,m}(z).
\label{eq:ASEP-rho1}  
\end{align}

Next, we have $\mathcal{N}^2=\sum_{n,m,n',m'} \langle \mathrm{n}_{n,m} \mathrm{n}_{n',m'}\rangle$, yielding the result that $\rho_\mathrm{tot}^2=(1/\mathcal{L}^2)\sum_{n,m,n',m'} \langle \mathrm{n}_{n,m} \mathrm{n}_{n'.m'}\rangle$. On the other hand, Eq.~\eqref{eq:ASEP-rho} gives
$\rho_\mathrm{tot}^2=(z^2 /\mathcal{L}^2)\sum_{n,m,n',m'} (\partial \ln F_{n,m}(z)/\partial z)(\partial \ln F_{n',m'}(z)/\partial z)$. Then, on comparison and using Eq.~\eqref{eq:ASEP-rho1}, we get 
\begin{align}
    \langle \mathrm{n}_{n,m} \mathrm{n}_{n',m'} \rangle=\rho_{n,m} \rho_{n',m'}= \langle \mathrm{n}_{n.m} \rangle \langle \mathrm{n}_{n',m'} \rangle.
    \label{eq:ASEP-15}
\end{align}

On the basis of the above, we obtain on using Eqs.~\eqref{eq:branch-again-0}-\eqref{eq:branch-again-1} and~\eqref{eq:backbone-again-0} that for the branch sites, we have
\numparts
\begin{align}
0 &= \alpha \rho_{n, L_{n}-1} - \beta  \rho_{n,L_{n}}- (\alpha - \beta) \rho_{n,L_{n}-1}~\rho_{n,L_{n}};\nonumber \\
&\qquad \qquad \qquad \qquad \qquad \qquad m = L_{n}~\text{and} ~\forall~ n=0,1,2,\ldots,N-1, \label{eq:1}\\
0&=  \alpha \rho_{n, m-1}+ \beta \rho_{n,m+1} - (\alpha + \beta) \rho_{n,m}+(\alpha - \beta) \rho_{n,m} \left[ \rho_{n,m+1}    - \rho_{n,m-1}\right]; \nonumber \\
&\qquad \qquad \qquad \qquad \qquad \qquad 0 < m < L_{n}~\text{and} ~\forall~ n=0,1,2,\ldots,N-1, 
\label{eq:ME_branch}
\end{align}
\endnumparts
while for the backbone sites, we have
\begin{align}
0&=[\alpha \rho_{n-1, 0} + \beta \rho_{n+1,0} +  \beta \rho_{n,1} - (2 \alpha + \beta) \rho_{n,0}] \nonumber \\
        &+ (\alpha - \beta) \rho_{n,0} \left [   \rho_{n+1,0}   + \rho_{n,1} - \rho_{n-1, 0} \right];\qquad 0 \le n \le N-1, \label{eq:backbone-1}
\end{align}
with the conditions $\rho_{N, 0}=\rho_{0, 0}$ and $\rho_{-1, 0}=\rho_{N-1, 0}$.

Equation~\eqref{eq:1} implies that $\alpha \rho_{n, L_{n}-1} - \beta  \rho_{n,L_{n}} = (\alpha - \beta) \rho_{n,L_{n}-1}~\rho_{n,L_{n}}$, yielding
\begin{align}
\rho_{n, L_{n}} =  \frac{\rho_{n,L_n-1}}{\frac{1}{f} + \left(1-\frac{1}{f}\right)\rho_{n,L_n-1}}.
	\label{eq:ME_BB_A}
\end{align}
Similarly, Eq.~\eqref{eq:ME_branch} yields 
\begin{align}
\alpha \rho_{n, m-1} + \beta \rho_{n,m+1} - (\alpha + \beta) \rho_{n,m} = (\alpha - \beta) \rho_{n,m} \left[ \rho_{n,m-1}   - \rho_{n,m+1}\right];\quad0 < m < L_{n}.
\label{eq:ME_BB_B}
\end{align}
Substituting $m=L_{n} -1$ in the above equation and using Eq.~\eqref{eq:ME_BB_A}, we get
\begin{align}
	\rho_{n, L_{n}-1} =  \frac{\rho_{n,L_n-2}}{\frac{1}{f} + \left(1-\frac{1}{f}\right)\rho_{n,L_n-2}}.
	\label{eq:ME_BB_B3}
\end{align}
The above equation relates the stationary-state densities between sites at distances $1$ and $2$ units from the reflecting end of the $n$-th branch. In this way, substituting successively for different values of $m$ in Eq.~\eqref{eq:ME_BB_B}, we obtain a relation between the stationary-state densities on two consecutive sites $(n,m)$ and $(n, m-1)$ on a branch, as  
\begin{align}
	\rho_{n, m} =  \frac{\rho_{n,m-1}}{\frac{1}{f} + \left(1-\frac{1}{f}\right)\rho_{n,m-1}};\qquad 1\leq m \leq L_n,
	\label{eq:ME_BB_m}
\end{align}
leading finally to
\begin{align}
	\rho_{n, m} = \left [ 1 + \frac{1}{f^m} \left ( \frac{1}{\rho_{n,0}}  - 1 \right ) \right ]^{-1}. 
	\label{eq:ME_BB_E}
\end{align}
The above equation relates the stationary-state density on any branch site to that on the corresponding backbone site. 

We now turn to the stationary-state densities on backbone sites. Equation~\eqref{eq:backbone-1} gives for $ 0 < n < N-1$ that
\begin{align}
&\left[\rho_{n-1, 0} + \frac{1}{f} \rho_{n+1,0} -  \left(2 + \frac{1}{f}\right) \rho_{n,0} \right]
        + \left(1 - \frac{1}{f}\right) \rho_{n,0}\left [   \rho_{n+1,0} - \rho_{n-1, 0} \right] \nonumber \\&+ \Big( \frac{1}{f} + \left(1 - \frac{1}{f}\right) \rho_{n,0} \Big) \rho_{n, 1} = 0.
		\label{eq:ME_BB0}
\end{align}    
Substituting for $\rho_{n, 1}$ from Eq.~\eqref{eq:ME_BB_m} in Eq.~\eqref{eq:ME_BB0}, we get
\begin{align}
    \left[\rho_{n-1, 0} + \frac{1}{f} \rho_{n+1,0} -  \left(1 + \frac{1}{f}\right) \rho_{n,0} \right]
    + \left(1 - \frac{1}{f}\right) \rho_{n,0}\left [   \rho_{n+1,0} - \rho_{n-1, 0} \right] = 0,
	\label{eq:ME_BB1}
\end{align}
which applies to all $0 \le n \le N-1$ with the conditions $\rho_{N, 0}=\rho_{0,0}$ and $\rho_{-1, 0}=\rho_{N-1, 0}$.

We have obtained a remarkable result: Eq.~\eqref{eq:ME_BB1} (i) is independent of $L_n$, and, moreover, (ii) is equivalent to that with $L_n=0$. The latter means more specifically that Eq.~\eqref{eq:ME_BB1} is mathematically equivalent to that for stationary-state single-site densities $\sigma_n^\mathrm{st};~n=0,1,\ldots,N-1$ for ASEP particles undergoing hopping to nearest-neighbor sites on a $1D$ periodic lattice of $N$ sites, with $\alpha$ and $\beta$ being the forward and the backward hopping rate. This aforementioned equivalence with respect to ASEP dynamics on a $1D$ lattice holds only for stationary-state single-site densities, and holds despite the fact that the underlying dynamics includes backbone and branch sites and involves hopping between them. The physical reason for this equivalence is that the average current in the branches is zero in the stationary state, as we show later in Section~\ref{subsec:drift}. We will later use this equivalence to obtain analytical results on the stationary-state transport properties of the ASEP particles on the comb. The stationary-state density equation being equivalent yields in both cases a uniform density: uniform ($\rho_{n,0}=\rho^\mathrm{st}~\forall~n$) over the backbone sites, and uniform ($={\sigma}^\mathrm{st}$) over the $1D$ periodic lattice. However, the total number of ASEP particles $\mathcal{N}$ being conserved in both cases, the normalization condition for the stationary-state density reads differently. Indeed, one has $\sum_{n=0}^{N-1} \sum_{m=0}^{L_n} \rho_{n,m} = \mathcal{N}$ for the comb, whereas  for the periodic lattice, one has instead that $\sum_{n=0}^{N-1} \sigma^{\mathrm{st}} = \mathcal{N}$. 

Using Eq.~\eqref{eq:ME_BB_E} in the normalization condition $\sum_{n=0}^{N-1} \sum_{m=0}^{L_n} \rho_{n,m} = \mathcal{N}$ gives 
\begin{align}
    \sum_{n=0}^{N-1} \sum_{m=0}^{L_n} \left[  1 + \frac{1}{f^m} \left ( \frac{1}{\rho^{\mathrm{st}}}  - 1 \right ) \right ]^{-1} = \mathcal{N}.
\label{eq:norm}
\end{align}
One now requires to solve the above algebraic equation numerically, the root of which yields the stationary-state uniform density $\rho^\mathrm{st}$ on the comb-backbone. Knowing $\rho^\mathrm{st}$, the stationary-state branch-site densities can be readily obtained from Eq.~\eqref{eq:ME_BB_E}. We thus note that although in the stationary state, the densities are uniform over the backbone, the same on the branches are not at all uniform. Equation~\eqref{eq:ME_BB_E} also shows how the stationary-state densities are distributed on a branch, being lowest on the corresponding backbone site and highest at the branch end point. Longer the branch, higher are the densities towards the branch end. Explicitly, consider two branches of lengths $L_{n_1}$ and $L_{n_2}$ attached to backbone sites $n_1$ and $n_2$, respectively, with $L_{n_1} > L_{n_2}$. Then, one can show on using Eq.~\eqref{eq:ME_BB_E} that $\rho_{n_1, L_{n_1}} > \rho_{n_2, L_{n_2}}$.

We now argue that $\rho^{\mathrm{st}}$ in Eq.~\eqref{eq:norm} does indeed represent a physical quantity, namely, the ASEP particle density, and as such satisfies $0 \leq \rho^{\mathrm{st}} \leq 1$. 
Let us say that instead, we have $\rho^{\mathrm{st}} \gae 1$, implying $ 1/\rho^{\rm{st}}= 1 - \epsilon$, with $ 0<\epsilon \ll 1 $; we will argue below that such a range of values for $\rho^{\mathrm{st}}$ is not allowed by Eq.~\eqref{eq:norm}. With $ 1/\rho^{\rm{st}}= 1 - \epsilon$, we have $ \sum_{m=0}^{L_n} \left[  1 + \frac{1}{f^m} \left ( \frac{1}{\rho^{\mathrm{st}}}  - 1 \right ) \right ]^{-1} > (L_n + 1) + \epsilon \frac{1 - f^{-(L_n+1)}}{1 - f^{-1}}$. Denoting the summation in Eq.~\eqref{eq:norm} by $S$, we then have $ S > \sum_{n=0}^{N-1} \left[ (L_n + 1) + \epsilon \frac{1 - f^{-(L_n+1)}}{1 - f^{-1}} \right]$. Given that $ f > 1 $ and $ \sum_{n=0}^{N-1} (L_n + 1) \geq \mathcal{N}$ (the total number of RC sites can be equal to or greater than the total number of particles), we obtain $ S > \mathcal{N}$, which contradicts the condition $S=\mathcal{N}$ given by Eq.~\eqref{eq:norm}. Let us now demonstrate that $\rho^{\mathrm{st}} \geq 0$.
Assume $\rho^{\mathrm{st}} =-\epsilon$, with $ 0<\epsilon \ll 1 $, which results in $\left[  1 + \frac{1}{f^m} \left ( \frac{1}{\rho^{\mathrm{st}}}  - 1 \right ) \right ]^{-1} \approx - \epsilon f^m < 0$, implying $S <0$, thus contradicting Eq.~\eqref{eq:norm} in which we have $\mathcal{N} > 0$.
We therefore conclude that $ 0\leq \rho^{\mathrm{st}} \leq 1$, and Eq.~\eqref{eq:ME_BB_E} then implies that $0 \leq \rho_{n,m}^{\mathrm{st}} \leq 1~\forall~n, m>0$.

On the basis of the foregoing, we conclude that the product measure in Eq.~\eqref{eq:product-measure-RC} represents a valid stationary state of the ASEP dynamics on the RC; the constant density $ \rho^{\mathrm{st}}$ on the RC backbone and the non-uniform density $\rho_{n,m>0}$ on the branches implied by the product-measure (see Eqs.~\eqref{eq:norm} and~\eqref{eq:ME_BB_E}) have values in the correct range. Combined with the fact that the ASEP dynamics on the RC is ergodic and hence has a unique stationary state, we conclude that Eq.~\eqref{eq:product-measure-RC} represents the exact stationary state of the dynamics. 

\subsection{Stationary-state transport properties}
\label{subsec:drift}
We now proceed to obtain the transport properties of our system in the stationary state. We are particularly interested in the drift velocity along the backbone, defined as the velocity, computed along the backbone, of a particle (any particle) performing the ASEP dynamics on the RC. One may operationally define this velocity thus: once the system has settled into the stationary state, locate the different ASEP particles on the lattice and measure for each the distance it covers along the backbone over a large time duration in a typical realization of the dynamics. Summing these distances and dividing by the product of the time duration and the total number of particles yield the drift velocity for the given dynamical realization; Averaging over dynamical realizations then yields the stationary-state drift velocity we are interested in.    

Let us denote by $ v^\mathrm{st}_\mathrm{{drift}}$ the stationary-state drift velocity. To proceed, consider the stationary-state-equivalent problem of the ASEP dynamics on a $1D$ periodic lattice that we discussed in the preceding subsection. Evidently, the average current across any bond $(i,i+1)$ will be the same for every bond in the stationary state, and will be given by $J^{\mathrm{st}}_\mathrm{{drift}}=(\alpha - \beta)\sigma^{\mathrm{st}} (1-\sigma^{\mathrm{st}})$, using the fact that the stationary state corresponds to a product measure and a uniform density over the lattice. We thus immediately obtain the stationary-state drift velocity in this equivalent ASEP problem as $v^\mathrm{st}_\mathrm{{drift}} = J^{\mathrm{st}}_\mathrm{{drift}}/\sigma^{\mathrm{st}} = (\alpha - \beta) \, (1- \sigma^{\mathrm{st}})$. Using the equivalence unveiled in the preceding subsection between the stationary-state single-site densities for the RC backbone and that for an ASEP on a $1D$ periodic lattice, we thus obtain for our problem on the random comb that the stationary-state current along the backbone equals $J^{\mathrm{st}}_\mathrm{{drift}}=(\alpha - \beta)\rho^{\mathrm{st}} (1-\rho^{\mathrm{st}})$. 

Before proceeding to obtain an expression for $v^\mathrm{st}_\mathrm{{drift}}$, we now demonstrate that in the stationary state, the average current in the branches is identically zero. Denoting by $J_{(n,m-1) \to (n,m)}^\mathrm{st}$ the average of the net current from the site $(n, m-1)$ to the site $(n, m)$ in the stationary state, we have   
\begin{align}
       J_{(n,m-1) \to (n,m)}^\mathrm{st} &=\alpha \, \rho_{n,m-1} (1- \rho_{n, m}) - \beta \, \rho_{n,m} (1- \rho_{n, m - 1}). 
\end{align}
Substituting for $\rho_{n, m}$ from Eq.~\eqref{eq:ME_BB_m}, one arrives at the result: $J_{(n,m-1) \to (n,m)}^\mathrm{st}  = 0,~\, \forall~\, m>0,n $. 

Now, we proceed to compute the drift velocity $v^\mathrm{st}_\mathrm{{drift}}$ for the random comb. Note that this drift velocity considers displacements only along the backbone in the direction of the field and is thus defined as follows. If $T_N$ be the mean time of traversal of an ASEP particle through an RC with backbone of $N$ sites, the drift velocity is defined as $v^\mathrm{st}_\mathrm{{drift}} = \lim_{N \to \infty} N/T_N$, where we have taken the lattice spacing to be unity. During its evolution, a particle may undergo multiple excursions into the branches before eventually traversing the entire comb. Since, as argued above, the branches do not contribute to the stationary current on the comb, the quantity $J^{\mathrm{st}}_\mathrm{{drift}}/\rho_\mathrm{tot}$ serves as an `effective' velocity on the RC, and $T_N$ is given by $T_N = \frac{\mathcal{L}}{(J^{\mathrm{st}}_\mathrm{{drift}}/\rho_\mathrm{tot})}$, 
yielding finally that
\begin{align}
 v^\mathrm{st}_\mathrm{{drift}}=p \frac{J^{\mathrm{st}}_\mathrm{{drift}}}{\rho_\mathrm{tot}}=(\alpha - \beta)\,p \, \frac{\rho^{\mathrm{st}}(1- \rho^{\mathrm{st}})}{\rho_{\mathrm{tot}}}= 2 W g \,p \, \frac{\rho^{\mathrm{st}}(1- \rho^{\mathrm{st}})}{\rho_{\mathrm{tot}}},  
 \label{eq:drift_vel}
\end{align}
where $p\equiv N/\mathcal{L}$ is the fraction of the total number of sites that constitute the backbone. Note that $\rho_{\mathrm{tot}}$ and $p$ are random variables varying between realizations of the RC.

\begin{figure}[!htbp]
	\centering
	\includegraphics[scale=0.55]{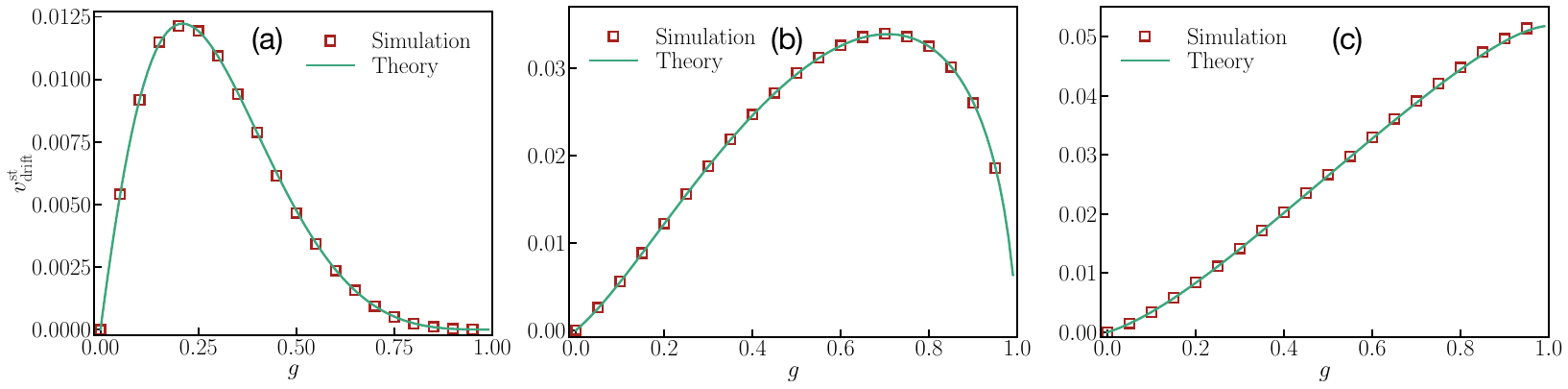}
	\caption{Shown in panels (a) -- (c) is the stationary-state drift velocity versus bias $g$ at various values of the total ASEP particle density $\rho_\mathrm{tot}$, as obtained from theory (Eq.~\eqref{eq:drift_vel},  continuous line) and numerical simulations (symbols), with number of backbone sites $N=100$  and for a typical disorder realization of the comb. The realization of the comb being fixed, varying $\rho_\mathrm{tot}$ is tantamount to varying the total number of particles $\mathcal{N}$ in the system. The branch lengths are sampled independently from the exponential distribution~\eqref{eq:exponential} with $M=20$ and $\xi=5$. The number of particles is $\mathcal{N}= 200$ [panel (a)], $ 400$ [panel (b)] and $450$ [panel (c)], whereas the total number of  sites on the comb is $513$. Numerics correspond to Monte Carlo simulations of the dynamics as detailed in~\ref{app1}; the displayed data correspond to the stationary state.}
	\label{fig:asep_result}
\end{figure}

\begin{figure}[!ht]
	\centering
	\includegraphics[scale=0.7]{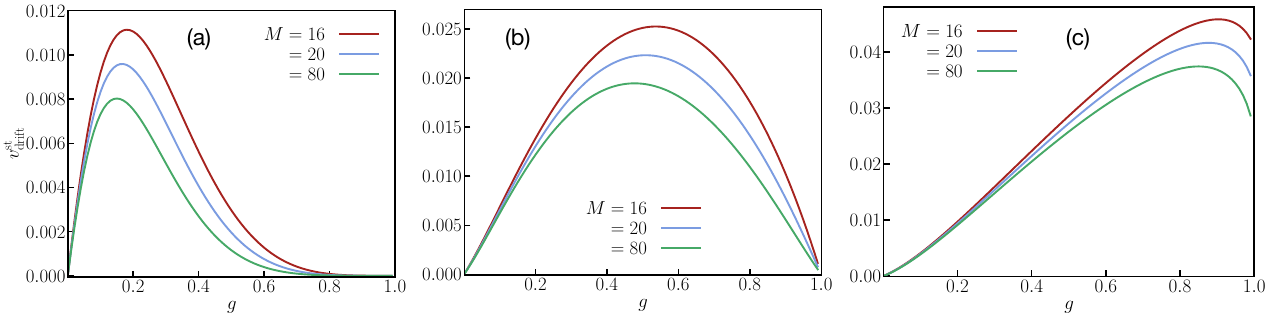}
	\caption{Panels (a) -- (c) show the behavior of stationary-state drift velocity with bias $g$ for three different ASEP particle density, $\rho_{\mathrm{tot}} = 0.35$, $0.70$, and $0.85$, respectively. Data are obtained from theory (Eq.~\eqref{eq:drift_vel}), with number of backbone sites $N=10^4$, and for a typical realization of the random comb. The branch lengths are sampled independently from the exponential distribution~\eqref{eq:exponential} with $\xi=5$ and $M=16, 20, 80$.}
	\label{fig:asep_drift_trapping}
\end{figure}

\begin{figure}[!ht]
	\centering
    \includegraphics[scale=0.5]{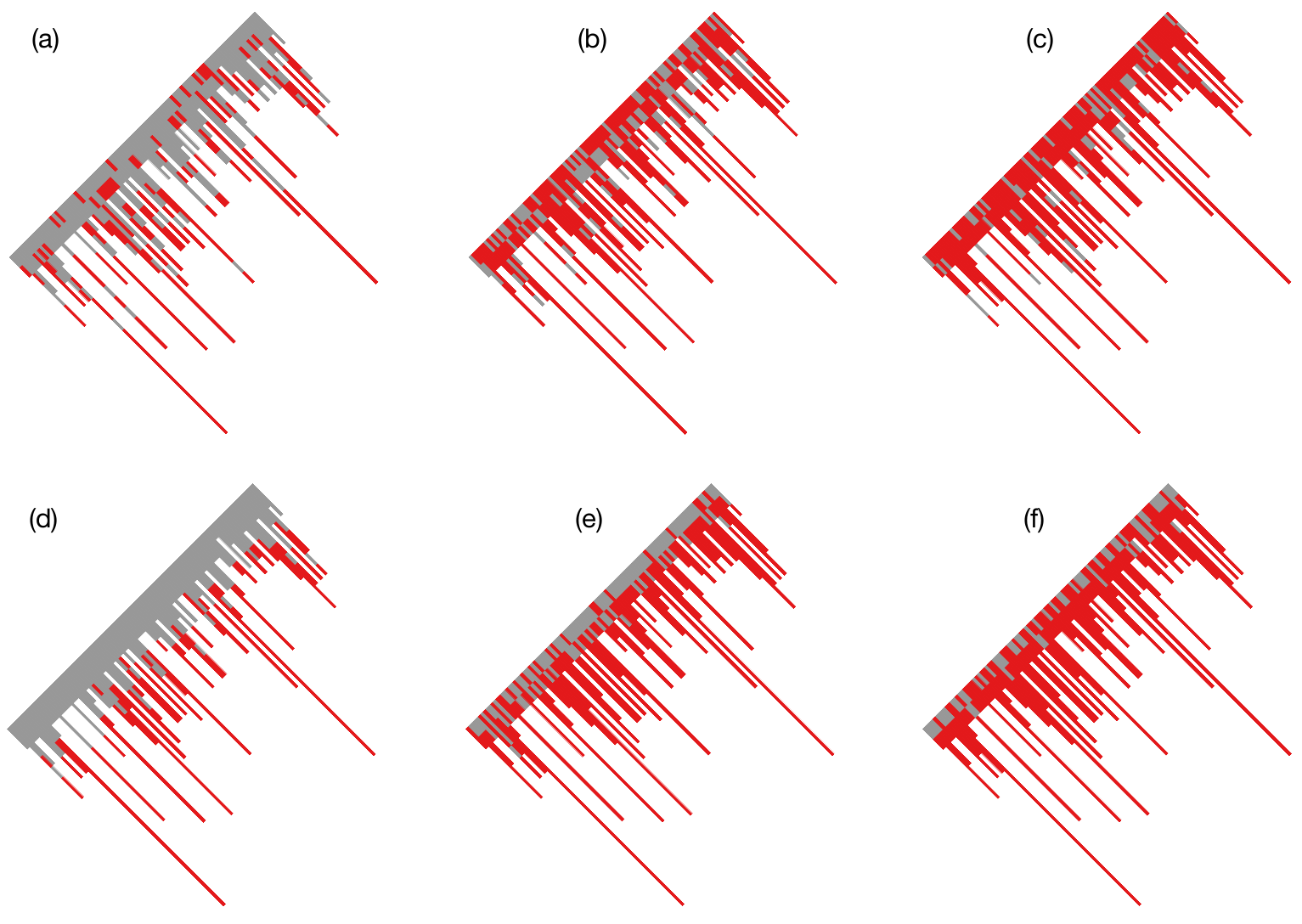}
	\caption{The pair of panels ((a),(d)), ((b),(e)) and ((c),(f)) show for the same values of $\rho_\mathrm{tot}$ (or, equivalently $\mathcal{N}$), $M$, $\xi$ as for Fig.~\ref{fig:asep_result}(a), (b), (c), respectively, a stationary-state (i.e., at long times $t=4\times 10^5$) snapshot of the random comb, with red representing occupied sites and gray denoting unoccupied ones, at a low bias $g=0.2$ and a high bias $g=0.8$, respectively.}
	\label{fig:asep_snapshot}
\end{figure}

We verify our result~\eqref{eq:drift_vel} with Monte Carlo simulation results at low as well as high ASEP particle density $\rho_{\mathrm{tot}}$, see Fig.~\ref{fig:asep_result}.  Details of the simulation procedure are given in~\ref{app1}. Note that in both simulations and displayed theoretical results in this paper, we take $W=1/2$, unless stated otherwise. In simulations for a fixed RC realization, note that varying particle density is tantamount to varying the total number $\mathcal{N}$ of particles in the system. Figure~\ref{fig:asep_drift_trapping}, panels (a) -- (c), show for three values of $\rho_\mathrm{tot}$, low, intermediate and high, our theoretical results on the variation of $v^\mathrm{st}_\mathrm{{drift}}$ with $g$; in each case, we have considered three values of $M$. Let us now discuss the displayed results. For fixed values of $N$ (taken to be very large, $N=10^4$), $M$, and $\xi$, the most evident feature seen in panels (a) -- (c) is the following: at a fixed $g$, as $\rho_\mathrm{tot}$ increases, so does the drift velocity. Consequently, the variation of $v^\mathrm{st}_\mathrm{{drift}}$ with $g$ becomes more and more monotonic as $\rho_\mathrm{tot}$ increases beyond a threshold value. Moreover, as $M$ becomes larger, this threshold value for $\rho_\mathrm{tot}$ also increases. For given values of $N$, $M$, $\xi$, as $\rho_\mathrm{tot} \to 0$, the behavior of the drift velocity versus $g$ is expected to be the one reported earlier for the case of non-interacting particles undergoing biased hopping on the RC~\cite{white1984field}, implying a non-motonic dependence and existence of a critical $g$ beyond which the drift velocity becomes zero. Non-monotonic variation in transport properties with respect to variation in field strength is also observed in other contexts, most notably, in models with kinetic constraints~\cite{PhysRevE.86.031112}, which translated to ASEP dynamics stipulates that hopping of the particle to an empty site takes place only when the particle has at least two empty neighbors before and after the move. Note that the ASEP dynamics on the RC has no such in-built kinetic constraint.

In order to understand physically the aforementioned features for Fig.~\ref{fig:asep_result}, panels (a) -- (c), we now do the following. For panel \ref{fig:asep_result}(a), we show in panels (a) and (d) of Fig.~\ref{fig:asep_snapshot}, a typical snapshot of the configuration of the system at late times (i.e., in the stationary state) and for same values of $\rho_\mathrm{tot}$ (or, equivalently, the total particle number $\mathcal{N}$), $M$, $\xi$ as in panel \ref{fig:asep_result}(a), but with respectively two different values of bias $g$, one low and one high. We see that at a given $g$, the particles undergoing biased hopping in the direction of the bias fill up the branches, which in our RC all run along the direction of the bias. Once the branches get filled up, particles especially towards the end of the branches find it difficult to move back to the backbone. This is owing to the combined effect of the requirement behind such a move to hop against the bias (which occurs with a smaller rate relative to hops in the direction of the bias) and the presence of hard-core exclusion. The longer the branch, the higher is the probability that more particles towards the end of the branch are trapped. Consequently, only a few particles on the backbone and near the intersection of the backbone with the branches are mobile; let us denote this region by $\mathcal{R}$. This effect gets pronounced with increase of $g$, when even fewer particles in region $\mathcal{R}$ are mobile. This is seen by comparing region $\mathcal{R}$ in panels (a) and (d), when one finds in the latter panel fewer red dots in region $\mathcal{R}$ compared to the former panel. Repeating the exercise performed above for panel \ref{fig:asep_result}(a) also for panels \ref{fig:asep_result}(b) and \ref{fig:asep_result}(c), we may arrive at a similar conclusion by comparing the snapshots in panels (b),(e) and (c),(f), of Fig.~\ref{fig:asep_snapshot} respectively. Of course, with increase of $\rho_\mathrm{tot}$, one has in region $\mathcal{R}$ more particles as one moves from panels (a) to (b) to (c) and from panels (d) to (e) to (f). 

On the basis of the above, we conclude that as $\rho_\mathrm{tot}$ increases, keeping $g$, $M$, $\xi$ fixed, the branches become less participatory in the dynamics, and it is only the mobile particles in region $\mathcal{R}$ (whose number increases with increasing $\rho_\mathrm{tot}$) that contribute to $v^\mathrm{st}_\mathrm{{drift}}$. These latter particles cannot get trapped, as the effects of branches are largely absent. Consequently, the variation of $v^\mathrm{st}_\mathrm{{drift}}$ with $g$ becomes more and more monotonic as $\rho_\mathrm{tot}$ increases beyond a threshold value. An estimate of the latter would be the density corresponding to all branch sites filled up, given by the ratio $(1 - N/\mathcal{L})$.

\section{Summary and conclusions}
\label{sec:summary}
In this work, we obtained the exact stationary-state static and dynamic properties of particles undergoing ASEP dynamics on an RC. Our results show that the stationary-state density is uniform along the backbone and nonuniform along the branches, decreasing monotonically from the free-end of a branch to its intersection with the backbone. On the other hand, the drift velocity of particles along the backbone when studied as a function of the bias strength exhibits a non-monotonic dependence, which remarkably becomes increasingly monotonic as one increases the ASEP particle density. This effect is a manifestation of an effective reduction of the branch lengths and a motion of the particles that takes place primarily along the backbone, owing to an intricate interplay between hard-core interactions and biased hopping. Introducing stochastic resetting\,~\cite{evans2020stochastic,gupta2022stochastic} in the dynamics is an interesting future direction worth taking up, an issue that was studied by us recently in the absence of any interaction between the particles and was shown to lead to nontrivial results of relevance~\cite{sarkar2022biased}. 

\section{Acknowledgements} This project is supported by the Deutsche Forschungsgemeinschaft (DFG, German Research Foundation) under Germany’s Excellence Strategy EXC 2181/1-390900948 (the Heidelberg STRUCTURES Excellence Cluster). M.S. also acknowledges support by the state of Baden-Württemberg through bwHPC cluster. S.G. acknowledges computational resources of the Department of Theoretical Physics, TIFR, assistance of Kapil Ghadiali and Ajay Salve, and financial support of Department of Atomic Energy, Government of India, under Project Identification No. RTI 4002. We are very grateful to the anonymous referee for pointing out to us the applicability of the pairwise balance condition to our system and for providing the reasoning presented in~\ref{app2}.

\appendix
\section{Explicit form of the master equation in Eq.~\eqref{eq:ME-RC}}
\label{app0}
Here, we provide the explicit form of the master equation in Eq.~\eqref{eq:ME-RC}: 

\begin{align}
    \frac{\partial P(\{\mathrm{n}_{n,m}\}, t)}{\partial t} &= \sum_{n=0}^{N-1}\sum_{m=1}^{L_n-1} \Bigg( \underbrace{\alpha P(\{\ldots, \underbrace{1,0}_{(n,m-1),(n,m)}, \ldots \}, t) + \beta P(\{\ldots, \underbrace{0,1}_{(n,m),(n,m+1)}, \ldots \}, t)}_{\text{Branch site contribution (influx)}} \nonumber\\
    &\quad - \underbrace{\alpha P(\{\ldots, \underbrace{1,0}_{(n,m),(n,m+1)}, \ldots \}, t) - \beta P(\{\ldots, \underbrace{0,1}_{(n,m-1),(n,m)}, \ldots \}, t) }_{\text{Branch site contribution (outflux)}}\Bigg) \nonumber\\
    &\quad + \sum_{n=0}^{N-1} \Bigg(\underbrace{ \alpha P(\{\ldots, \underbrace{1,0}_{(n,L_n-1),(n,L_n)}, \ldots \}, t) - \beta P(\{\ldots, \underbrace{0,1}_{(n,L_n-1),(n,L_n)}, \ldots \}, t)}_{\text{Branch end point contribution}} \Bigg) \nonumber\\
    &\quad + \sum_{n=1}^{N-2} \Bigg( \underbrace{\alpha P(\{\ldots, \underbrace{1,0}_{(n-1,0),(n,0)}, \ldots \}, t) + \beta P(\{\ldots, \underbrace{0,1}_{(n,0),(n+1,0)}, \ldots \}, t)}_{\text{Backbone site contribution (influx)}} \nonumber\\
    &\quad - \underbrace{\alpha P(\{\ldots, \underbrace{1,0}_{(n,0),(n+1,0)}, \ldots \}, t) - \beta P(\{\ldots, \underbrace{0,1}_{(n-1,0),(n,0)}, \ldots \}, t)}_{\text{Backbone site contribution (outflux)}} \nonumber\\
    &\quad - \underbrace{\alpha P(\{\ldots, \underbrace{1}_{(n,0)}, \ldots, \underbrace{0}_{(n,1)}, \ldots \}, t) + \beta P(\{\ldots, \underbrace{0}_{(n,0)}, \ldots, \underbrace{1}_{(n,1)}, \ldots \}, t)}_{\text{Backbone to branch contribution}} \Bigg) \nonumber\\
    &\quad + \Bigg( \alpha P(\{\underbrace{0}_{(0,0)}, \ldots, \underbrace{1}_{(N-1,0)}, \ldots \}, t) + \beta P(\{\underbrace{0,1}_{(0,0),(1,0)}, \ldots \}, t) \nonumber\\
    &\quad - \alpha P(\{\underbrace{1,0}_{(0,0),(1,0)}, \ldots \}, t) - \beta P(\{\underbrace{1}_{(0,0)}, \ldots, \underbrace{0}_{(N-1,0)}, \ldots \}, t) \nonumber\\
    &\quad - \underbrace{\alpha P(\{\underbrace{1}_{(0,0)}, \ldots, \underbrace{0}_{(0,1)} \}, t) + \beta P(\{\underbrace{0}_{(0,0)}, \ldots, \underbrace{1}_{(0,1)}, \ldots \}, t)}_{\text{Backbone site boundary contribution}} \Bigg) \nonumber\\
    &\quad + \Bigg( \alpha P(\{\ldots, \underbrace{1,0}_{(N-2,0), (N-1, 0)}, \ldots  \}, t) + \beta P(\{\underbrace{1}_{(0,0)}, \ldots, \underbrace{0}_{(N-1,0)}, \ldots \}, t) \nonumber\\
    &\quad - \alpha P(\{\underbrace{0}_{(0,0)}, \ldots, \underbrace{1}_{(N-1,0)}, \ldots \}, t) - \beta P(\{\ldots, \underbrace{0,1}_{(N-2,0), (N-1, 0)}, \ldots \}, t) \nonumber\\
    &\quad - \underbrace{\alpha P(\{\ldots, \underbrace{1,0}_{(N-1,0), (N-1,1)}, \ldots \}, t) + \beta P(\{\ldots, \underbrace{0,1}_{(N-1,0), (N-1,1)}, \ldots \}, t)}_{\text{Backbone site boundary contribution}} \Bigg).
    \label{eq:app0-ME-RC}
\end{align}

\section{Stationary-state measure of configurations on the RC}
\label{app2}
Consider first the ASEP dynamics on an RC branch, say, the $k$-th branch with length $L_k$. The average current in the branches is zero in the stationary state, and consequently, one may model the dynamics on the branch as an equilibrium dynamics that satisfies the condition of detailed balance and which leads to an equilibrium distribution of particles on the branch to be proportional to $e^{-\mathcal{H}_k/T}$, with $T$ a parameter, and with the Hamiltonian given by 
\begin{align}
\mathcal{H}_k(\{\mathrm{n}_j\}_{0\le j \le L_k}) = -E_0 \sum_{j=0}^{L_k} j \mathrm{n}_j.
\end{align}
Here, the site $j=0$ is shared by both the branch and the backbone, $C_k \equiv \{\mathrm{n}_j\}_{0\le j \le L_k}$ is a configuration of particles, and where $E_0$ is a constant. While considering dynamical moves for the site $j=0$, we consider moves only within the branch (in the later part of this appendix, we will consider particle moves between the branches and the backbone). The condition of detailed balance implies that for a transition from a configuration $C_k=\{...1,0,..\}$ to another configuration $C_k'= \{...0,1...\}$, the corresponding stationary probability $P(C_k)$ satisfies
\begin{align}
P(C_k) W(C_k \to C'_k) = P(C'_k) W(C'_k \to C_k).
\end{align}
Noting that $W(C_k \to C'_k)=W(1+g)$ and $W(C'_k \to C_k)=W(1-g)$ and that $P(C_k) \propto e^{-\mathcal{H}_k(C_k)/T}$, we get on using the above equation that 
\begin{align}
\frac{E_0}{T} &=\frac{1}{L_g};~~L_g \equiv 1/ \ln\left( \frac{1+g}{1-g} \right).
\label{app2-eq1}
\end{align}
The above equation fixes the value of the parameter $T$ for a given value of $g$.

Note that the ASEP dynamics on the RC conserves the total number of particles. Since the number of particles on the branch fluctuates as a function of time, we may invoke a grand canonical ensemble description involving a fugacity $z$ that naturally allows this number to fluctuate but fixes the average number of particles on the branch through the choice of the value of $z$, as is standard in equilibrium statistical mechanics. For the $k$-th branch, the grand canonical partition function reads as
\begin{align}
Z(z,g) =\sum_{\mathcal{N}_k=0}^\infty z^{\mathcal{N}_k}\sum_{\{\mathrm{n}_j\}_{0 \le j \le L_k}}e^{-\mathcal{H}_k(\{\mathrm{n}_j\})/T}\delta_{\sum_{j=0}^{L_k}\mathrm{n}_j,\mathcal{N}_k}=\prod_{j=0}^{L_k} \left( 1+ z e^{j/L_g} \right).
\end{align}

\begin{figure}[!ht]
\centering	\includegraphics[scale=0.5]{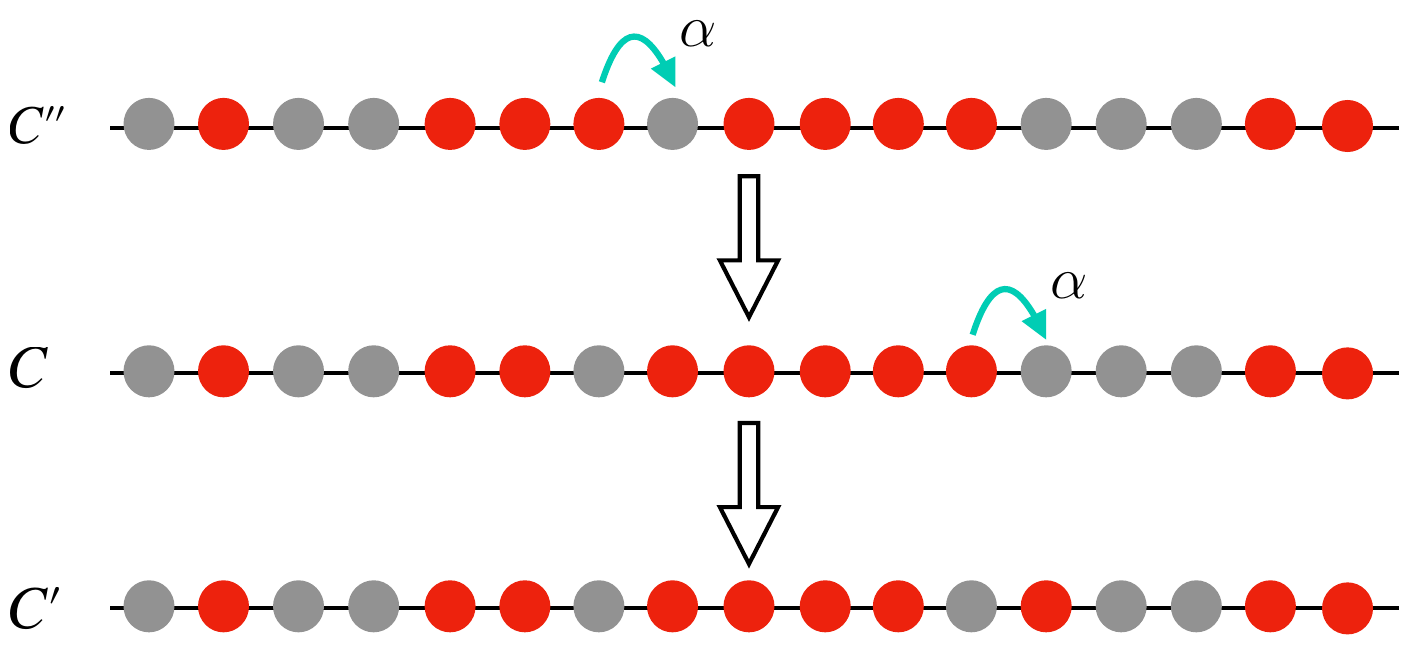}
\caption{For the ASEP dynamics on a $1D$ periodic lattice, the figure shows a schematic view of three representative configurations $C,C',C''$ such that the pairwise balance condition is satisfied.}
\label{fig:BB_ss_measure}
\end{figure}

Let us now discuss the dynamics on the backbone. Considering particle moves only within the backbone, one has the usual ASEP dynamics on a $1D$ periodic lattice. In this case, the stationary state satisfies the condition of pairwise balance: for any configuration $C$, the probability current to a configuration $C'$ is balanced by the same from another configuration $C''\ne C'$ into $C$. Thus, for every $C$, one needs to find a unique pair $(C',C'')$ such that the condition is satisfied. To this end, one defines for a given configuration a `particle cluster' as the maximal set of consecutive particles bounded by two empty sites. Referring to  Fig.~\ref{fig:BB_ss_measure}, in which $C'$ is obtained by moving one site ahead the rightmost particle in a particle cluster in $C$, we see that $C''$ is constructed by moving one site backward the leftmost particle in the same particle cluster in $C$. It is thus evident that for every $C$, a unique pair $(C', C'')$ can be identified such that the pairwise balance condition is satisfied. The pairwise balance condition then implies $P(C)=P(C')=P(C'')$, i.e., all configurations are equally likely in the stationary state. In the grand canonical ensemble, the probability of a configuration with a certain number $\mathcal{M}$ of particles on the backbone is then proportional to $y^{\mathcal{M}}$, where $y$ is the fugacity associated with the backbone.

We now proceed to consider particle moves between the backbone and the branches. Consider first the case of an RC with a single branch. The number of particles in both the branch and the backbone can change in course of dynamical evolution. In terms of the fugacities $z$ and $y$ introduced above, if $\mathcal{N}_b$ is the number of particles on the branch out of the total of $\mathcal{N}$ particles on the RC, the stationary-state probability of an RC configuration is given by
\begin{align}
P(\{\mathcal{N}_b\},\{\mathcal{N} - \mathcal{N}_b\}) = \frac{y^{\mathcal{N} - \mathcal{N}_b} z^{\mathcal{N}_b} e^{- \mathcal{H}(\{\mathcal{N}_b\})/T}}{\mathcal{Z}},
\end{align}
where $\{\mathcal{N}_b\}$ is a particle configuration on the branch consisting of $\mathcal{N}_b$ particles, $\{\mathcal{N} - \mathcal{N}_b\}$ is a particle configuration on the backbone consisting of $\mathcal{N} - \mathcal{N}_b$ particles, and $\mathcal{Z}$ is the partition function or the normalization constant. Writing then the detailed balance condition between two configurations 
$C \equiv (\{\mathcal{N}_b+1\}, \{\mathcal{N} - \mathcal{N}_b -1\})$ and $C' \equiv (\{\mathcal{N}_b\}, \{\mathcal{N} - \mathcal{N}_b\})$, 
\begin{align}
P(C) W(1-g) &= P(C') W(1+g), 
\end{align}
one obtains 
\begin{align}
(z/y) e^{E_0/T} = \frac{1+g}{1-g},
\end{align}
which on using Eq.~\eqref{app2-eq1} leads to the result that
\begin{align}
z = y.
\end{align}
Thus, the condition of detailed balance holds even for particle moves between the backbone and the branch provided the corresponding fugacities are equal. The above line of reasoning can be generalized to a comb with any number of branches, by considering a branch $\leftrightarrow$ backbone particle exchange for one branch at a time, while keeping the occupancies fixed in all the other branches.

\section{Details of Monte Carlo simulations}
\label{app1}
Here, we discuss a Monte Carlo simulation algorithm to simulate the ASEP dynamics on a given realization of the random comb. We first need to generate the comb. This is done by deciding on the number of backbone sites (the quantity $N$) and the branch-length cut-off $M$, and then drawing independently for each backbone site the corresponding branch length $L_{n}$ from the exponential distribution $\mathcal{P}_L$ in Eq.~\eqref{eq:exponential}.
We next choose specific values of the total number of ASEP particles $\mathcal{N} (\le \mathcal{L})$, the bias $g$, and the parameter $W$. As mentioned in the main text, the lattice spacing is taken to be unity. As representative choices to perform our simulations, we take $N=100$, $M=20$, and $W=1/2$, unless stated otherwise.  

 A typical simulation involves initializing the dynamics at time $t=0$ with $\mathcal{N}$ particles distributed randomly on various RC-sites $(n,m)$, and letting them perform the ASEP dynamics with chosen value of the time step $\mathrm{d}t$. Given the position of a particle at time $t$, in the ensuing infinitesimal time interval $[t,t+\mathrm{d}t]$, the position of the particle is updated as follows (provided of course that when the attempt of a particle to move away from its current location is accepted, the particle can actually hop into the destination site, in case the latter is unoccupied):
\begin{enumerate}
    \item If at time $t$ the particle is on a branch site that is not the end site of the branch, it attempts to move along the branch with equal probability of $1/3$ in either along or opposite to the direction of the bias, while it attempts to stay put with probability $1/3$. The move in the direction of the bias is actually accepted with probability $(3/2)(1+g) \mathrm{d}t$, while the move opposite to the direction of the bias is accepted with probability $(3/2)(1-g) \mathrm{d}t$.

    \item If at time $t$ the particle is on the end site (the reflecting end) of the branch, it attempts to move along the branch and in the direction opposite to the bias with probability $1/3$, while it attempts to stay put with probability $2/3$.  The move opposite to the direction of the bias is successful with probability $(3/2)(1-g) \mathrm{d}t$.

    \item If at time $t$ the particle is on a backbone site with no branch attached, it decides to move along the backbone either along or opposite to the direction of the bias with equal probability of $1/3$, while it stays put with probability $1/3$.  The moves in the direction of and opposite to the direction of the bias are accepted respectively with probabilities $(3/2)(1+g)\mathrm{d}t$ and $(3/2)(1-g)\mathrm{d}t$.

    \item If at time $t$ the particle is on a backbone site with a branch attached to it, it decides to move along the backbone either along or opposite to the direction of the bias with equal probability of $1/3$, while it decides to move into the attached branch site with probability $1/3$.  The moves in the direction of and opposite to the direction of the bias are accepted respectively with probabilities $(3/2)(1+g)\mathrm{d}t$ and $(3/2)(1-g)\mathrm{d}t$.  The move to the branch site is accepted with probability $(3/2)(1+g)\mathrm{d}t$. 
\end{enumerate}

One Monte Carlo time step of the dynamics corresponds to choosing for $\mathcal{N}$ number of times the ASEP particles at random, and updating their current location following the above-mentioned rules. We keep evolving the dynamics for a long time until it settles into a stationary state, and measure the stationary-state drift velocity in the following way.

In the stationary state, all the $\mathcal{N}$ particles are tracked for a long observation time $T$. We compute first the velocity of the individual particles, ($v[i];\, i=1, 2,\dots, \mathcal{N}$), along the backbone and in the direction of the bias as follows:
	\begin{align}
	\fl	v[i] =
      \frac{\substack{\text{Net displacement of the $i$-th particle on the backbone} \\
      \text{and in the direction of the bias in time}~T}}{T}. 
	\end{align}
	Finally, the stationary-state drift velocity is computed as
	\begin{align}
		v^\mathrm{st}_\mathrm{{drift}} = \frac{1}{\mathcal{N}} \sum_{i=1}^{\mathcal{N}} v[i].
	\end{align}

\newcommand{\newblock}{}
\bibliographystyle{unsrtnat}
\bibliography{References_mrinal}

\end{document}